\documentclass[apj]{emulateapj}

\def\gtorder{\mathrel{\raise.3ex\hbox{$>$}\mkern-14mu
             \lower0.6ex\hbox{$\sim$}}}
\def\ltorder{\mathrel{\raise.3ex\hbox{$<$}\mkern-14mu
             \lower0.6ex\hbox{$\sim$}}}

\slugcomment{Draft of \today}

\shorttitle{Radio transients}

\shortauthors{Frail et al.}

\begin{document}

\title{A Revised View of the Transient Radio Sky}

\author{
D.~A.~Frail\altaffilmark{1},
 S.~R.~Kulkarni\altaffilmark{2},
 E.~O.~Ofek\altaffilmark{2}$^,$\altaffilmark{3},
 G. C. Bower\altaffilmark{4} 
\&\
 E.  Nakar\altaffilmark{5}
}

\altaffiltext{1}{National Radio Astronomy Observatory, P.O. Box O,
  Socorro, NM 87801}
\altaffiltext{2}{Caltech Optical Observatories 249-17,
 California Institute of Technology, Pasadena, CA 91125, USA}
\altaffiltext{3}{Benoziyo Center for Astrophysics, 
Faculty of Physics,
The Weizmann Institute for Science, Rehovot 76100, Israel}
\altaffiltext{4}{Astronomy Department, University of California,
  Berkeley, 601 Campbell Hall \#3411, Berkeley, CA 94720-3411, USA}
\altaffiltext{5}{Raymond and Beverley Sackler School of Physics \&
  Astronomy, Tel Aviv University, Tel Aviv 69978, Israel}

\begin{abstract}

%

  We report on a re-analysis of archival data from the Very Large
  Array for a sample of ten long duration radio transients reported by
  Bower and others.  These transients have an implied all-sky rate
  that would make them the most common radio transient in the sky and
  yet most have no quiescent counterparts at other wavelengths and
  therefore no known progenitor (other than Galactic neutron stars).
  We find that more than half of these transients are due to rare data
  artifacts.  The remaining sources have lower signal-to-noise ratio
  (SNR) than initially reported by 1 to $1.5\,\sigma$.  This lowering
  of SNR matters greatly since the sources are at the threshold. We
  are unable to decisively account for the differences.
  By two orthogonal criteria one source appears to be a good
  detection.  Thus the rate of long duration radio transients without
  optical counterparts is, at best, comparable to that of the class of
  recently discovered Swift\,J1644+57 nuclear radio transients. We
  revisit the known and expected classes of long duration radio
  transients and conclude that the dynamic radio sky remains a rich
  area for further exploration.  Informed by the experience of past
  searches for radio transients, we suggest that future surveys pay
  closer attention to rare data errors and ensure that a wealth of
  sensitive multi-wavelength data be available in advance of the radio
  observations and that the radio searches should have assured
  follow-up resources.

\end{abstract}


\section{Introduction}\label{sec:Introduction}

A century ago, the study of variable stars was a leading area of
astronomy.  With the increasing availability of large format optical
detectors and inexpensive high speed computing this sub-field, as
witnessed by projects such as ASAS, OGLE, Catalina Sky Survey, the
Palomar Transient Factory, PanSTARSS and SkyMapper, is making a
come-back. Radio astronomy appears to be poised for a similar growth.
At meter wavelengths, commercially available signal processing chips
make it feasible to image the entire primary beam of a dipole or a
cluster of dipoles. These technological innovations lie at the heart
of LOFAR \citep{rbf03}, MWA \citep{lcm+09} and LWA \citep{ecc+09}.  At
centimeter wavelengths the ``large number, small diameter'' (LNSD)
array approach (made possible by inexpensive signal processing,
advances in commercial RF technology, innovative ideas in the design
of small diameter telescopes and phased array focal planes) has now
been demonstrated to be a cost effective method of building high speed
mapping machines \citep{Welch09,Dewdney09,Jonas09,ovc09}.  The LNSD
approach has motivated a new generation of radio facilities:
Apertif/WSRT \citep{ovc09}; MeerKAT \citep{bbjf09}; ASKAP
\citep{jtb+08}.

\begin{deluxetable*}{lllllll}[hbtp]
\tablecolumns{7}
\tabletypesize{\scriptsize} \tablewidth{0pt} \tablecaption{Long Duration Transient Populations} 
\tablehead{ \colhead{Class}   & \colhead{Rise}   & \colhead{Decay}   & \colhead{D}  & \colhead{Host} & \colhead{Rate}
& \colhead{Ref.}\\
& yr & yr &  & mag & deg$^{-2}$ & \cr
}
\startdata
Type II SNe       & 0.1-1  & 10    & 100\,Mpc   & 16   & 0.04 & \citet{gop+06}\\
Type Ib/c SNe     & 0.1    & 0.3   & 50\,Mpc    & 14.5 & $5\times 10^{-6}$ & \citet{bkfs03}\\
SN1998bw-like     & 0.1    & 0.1   & 300\,Mpc   & 18.4 & $3 \times 10^{-4}$ & \citet{skn+06} \\
Sw\,J1644+57-like & 0.1    & 1     & $z\sim1.8$ & 21.7 & 0.1 & \citet{zbs+11}\\
Orphan Afterglows & 1      & 1     & 1\,Gpc     & 21.0 & $10^{-2}$ & \citet{lowg02} \\
NS-NS mergers     & 0.1--1 & 0.1--3& 800\,Mpc   & 20.5 & $5 \times 10^{-3}$ & \citet{np11} \\
\enddata
\tablecomments
{
  Detectability distance and rates have been calculated assuming a
  single snapshot at a flux density threshold of 0.3\,mJy.  See
  section \ref{sec:logNlogS} for details.  $D$ is the distance at
  which the typical transient will have a specific flux of 0.3\,mJy.
  Host is the apparent magnitude of a galaxy with $-19$ absolute
  magnitude at distance $D$.  
}
\label{tab:ListOfTrans} 
\end{deluxetable*}

We divide radio transients into four categories based on two
attributes. The first is the duration of the basic phenomenon
(shorter than or greater than a few seconds).  The second is their
location (within the Galaxy or extra-galactic). Roughly speaking
the duration maps to coherent versus incoherent emission and the
location to repeated versus cataclysmic events.

Pulsars and related phenomenon (giant pulses, nulling pulsars,
erratic pulsars, rotating radio transients, and magnetars) are the
dominant category of short duration radio transients at meter and
centimeter wavelengths. There are no secure examples of short
duration radio transients that are located beyond the local Group.
Flare stars and associated phenomena are prime examples of long
duration radio transients of Galactic origin.

The focus of this paper is long duration transients of extra-galactic
origin. Known examples in this group are supernovae \citep{wps+10}
and gamma-ray burst afterglows \citep{grf09}. In both cases, the
radio emission arises as the fast moving debris interacts with the
circumstellar matter. In Table~\ref{tab:ListOfTrans} we summarize
the areal density of radio-emitting supernovae (including the
sub-classes) and GRB afterglows. Note the areal density of ``live
transients''  (transients present  at any given instant of time)
of both supernovae and GRB afterglows is less than 0.05 per square
degree.

In 2007, Bower et al.\nocite{bsb+07} [hereafter, B07] reported on the
analysis of a single field observed every week as a part of the Very
Large Array (VLA) calibration program. The observations were conducted
at 4.8\,GHz and 8.4\,GHz and lasted 22 years.  The 944 epochs and the
weekly cadence makes this data set a most valuable set to probe the
decimeter band for long duration transients at the sub-milliJansky
level.  These authors reported the discovery of eight transients found
in only one epoch (hereafter ``single-epoch''; duration, 20\,minutes
$<t_{\rm dur}<$ 1\,week) and two transients found in rolling 2-month
searches (hereafter, ``multi-epoch'' sources).

Deep observations towards these sources were undertaken at optical,
near-IR and X-ray bands.  The most remarkable feature of the B07
sources is an absence of optical and near-IR counterparts, despite
deep searches (B07; \citealt{obg+10}).  As noted by \citet{obg+10} all
extra-galactic transients (regardless of the band at which the
transient was discovered) have detectable optical counterparts,
namely, their host galaxies. Furthermore, remarkably the areal density
of transients live at any given time was estimated to be
1.5\,deg$^{-2}$ ($S>0.37\,$mJy).  This density exceeds that of all
other known radio transient source populations by an order of
magnitude (or more); see Table~\ref{tab:ListOfTrans}.  \cite{obg+10}
thus argued that the absence of an optical counterpart means that B07
transients have to be repeating sources of Galactic origin, and
proposed that B07 transients are old neutron stars (which naturally
satisfy the requirement of being optically almost invisible).

Given that the search for transient and strong variables is one of the
primary motivators for the next generation radio facilities (described
earlier) it is important to critically investigate the B07 transients
since this class nominally dominates over all other known classes of
radio transients (see Table~\ref{tab:ListOfTrans}).  To this end, here
we report on a re-analysis of the original data of B07
(\S\ref{sec:ObsReana}, \S\ref{sec:Findings}) and revisit the transient
reported by \citealt{ofb+11} (\S\ref{sec:OtherSurveys}).  In
\S\ref{sec:logNlogS} we present an update of the expected rate of
radio transients.  In \S\ref{sec:FutureSurveys} we discuss these rates
in relation to future synoptic radio imaging surveys and conclude.

\section{Observations and Re-analysis}
\label{sec:ObsReana}

The data used by B07 arose from a calibrator program carried out
during the period 1983--2005. All observations were made in the
standard continuum mode with 100\,MHz of total bandwidth in each
of two adjacent 50-MHz bands (IFs) at center frequencies of 5\,GHz
and 8.4\,GHz and in both hands of circular polarization. See B07
for more details about the full data-set.

For the re-analysis we selected, from the archive, only the raw data
relevant to the transients reported in B07. This means the eight
epochs from which the single-epoch transients were first found and the
3+8 data sets from which the two multi-epoch transients were found.
Data were taken at other radio frequencies in about half of the cases.
Some details of the single epoch and two-month transients can be found
in Table~\ref{tab:ListOfFields}).

For the re-analysis we used AIPS\footnote{http://www.aips.nrao.edu/}
\citep{Greisen03}.  The data reduction and imaging followed the same
path used by B07 with a small exception.  B07 employed AIPS for the
flagging and analysis of the single-epoch transients, and used the
{\it Miriad} package \citep{stw95}  for imaging the two-month averages.
We endeavored to make the calibration and the flagging of UV data
(AIPS task \texttt{TVFLG}) data for each epoch as uniform way.
Following these steps we ran each raw visibility data set through the
VLA pipeline (\texttt{VLARUN}).

No flux density calibrator was observed during any epoch of these
test observations. Following B07, the flux density of the phase
calibrator (B1803+784) was fixed to be 2.2\,Jy (5\,GHz) and 2.8\,Jy
(8.4\,GHz). For those epochs with 22\,GHz and 1.4\,GHz observations
the flux density of the phase calibrator was taken to be 3\,Jy and
2\,Jy, respectively. It is evident from the strong variations in
the radio light curves for B1803+784\footnote{U. Michigan Radio
Astronomy Observatory database} that these mean values are only
approximate.  Our reinvestigation confirm  that at least during the
period 1981--1999 the variation was less than 15\%.  Fortunately,
an accurate flux density scale is not crucial for our analysis since
we report results in terms of the signal-to-noise.

Following B07 we applied a Gaussian weighting to the visibility data
in order to limit the effects of bandwidth smearing.  This was done by
applying a 150-k$\lambda$ taper to all visibility data prior to
imaging (\texttt{IMAGR}). For each field we required that a source be
present in both frequency bands (IFs) with similar flux densities and
with similar positions. Images were made with extra large
fields-of-view.  The wide field-of-view is necessary to reduce the
effect of side-lobes that can mimic sources in narrow fields.  These
final analysis images had a size of about 40-arcmin at 5\,GHz data and
27-arcmin at 8.4\,GHz.  For guidance, the full-width at half power for
VLA antennas is 45-arcmin/$\nu$(GHz), or 9.3-arcmin at 5\,GHz and
4.3-arcmin for 8.4\,GHz. Measurements of the VLA beam power response
beyond the first null are given in \citet{cp10}, while the polynomial
coefficients needed to correct for the primary beam attenuation can be
found in the AIPS task \texttt{PBCOR}.  We do note that these
corrections are uncertain for large angular distances from the beam
center.  All data were taken in the B1950 coordinate system.  We
stayed in the B1950 system throughout calibration and imaging.

\begin{deluxetable*}{lllllll}
\tablecolumns{5}
\tabletypesize{\scriptsize}
\tablewidth{0pt}
\tablecaption{Single Epoch and Two-Month Transient Candidates}
\tablehead{
\colhead{Name}   &
\colhead{Type} &
\colhead{Freq.}   &
\colhead{FWHM/2} & 
\colhead{$\Delta\theta$}  &
\colhead{Beam}  &
\colhead{}\\
\colhead{Candidate}   &
\colhead{} &
\colhead{(GHz)}   &
\colhead{(arcmin)} & 
\colhead{(arcmin)}  &
\colhead{(arcsec)}  &
\colhead{Notes}
}
\startdata
RT\,19840502 & SE &  4.9 & 4.6 & 0.22& 6.0  & Phase center artifact.\\
RT\,19840613 & SE & 4.9 & 4.6 & 3.3 &  5.7 & Side-lobe of bright source.\\
RT\,19860115 & SE & 4.9 & 4.6 & 1.3 & 14.8 & Side-lobe of bright source.\\
RT\,19860122 & SE & 4.9 & 4.6 & 4.6 & 14.5 & Artifact. Lower IF is bad.\\
RT\,19870422 & 2M & 4.9 & 4.6 & 8.0 & 12.8 & Artifact. Bad pointing.\\
RT\,20010331 & 2M & 8.5 & 2.6 & 4.4 & 1.5  &  No detection (see \S\ref{sec:RT20010331}).\\
\\
RT\,19920826 & SE &  4.9 & 4.6 & 2.0 & 21.2 &{\bf D}\ SNR: {\it 6.4}; \texttt{5.8}.\\
RT\,19970205 & SE & 8.5 & 2.6 & 4.4 & 1.4  & {\bf B}\  SNR: {\it 7.7}; \texttt{5.7} (CA)\\
RT\,19970528 & SE & 4.9 & 4.6 & 6.8 & 3.9  & {\bf CnB}\ SNR: {\it 7.5}; \texttt{5.6}. (CA)\\
RT\,19990504 & SE & 4.9 & 4.6 & 8.9 & 18.9 &{\bf D}\ SNR: {\it 7.3}; \texttt{5.7}.\\
\\
RT\,19970528 & SE & 8.5 & 2.6 & 6.8 & 1.3  & {\bf CnB}\  No detection (BA)\\
RT\,19990504 & SE & 8.5 & 2.6 & 8.9 &  8.3 & {\bf D}\  No detection (BA)

\enddata

\tablecomments{Starting from the left the columns are as follows.  The
  name of the transient as RT\,YYYYMMDD where YYYY is the UT year, MM
  is the month index and DD is the day number at which the transient
  was first detected; the type of transient: single epoch (SE) or two
  month average (2M); the center frequency in GHz; one half of the
  full-width at half-maximum of the primary response beam in arc
  minutes; the offset of the transient from the phase center, also in
  arc minutes; the beam size in arc seconds computed as geometric mean
  of the major and minor axes;.  The last column report the array
  configuration, two SNRs (for sources detected and reported as such
  in B07) and some comments.  The left SNR (in {\it italics}) are the
  SNR from B07 and the right SNR (in \texttt{typewriter} font) are SNR
  resulting from the work presented here (see \S\ref{sec:Findings} for
  details).  The comments are as follows: CA for severe Chromatic
  Aberration and BA for severe Beam Attenuation.  The first group of
  transients are either artifacts or clearly very low SNR.  Following
  that (separated by a line) are four transients which are threshold
  sources. The last group is composed of higher frequency observations
  of two of these threshold sources and given the strong beam
  attenuation there was no expectation of detection. The entries were
  made merely for completeness.}
\label{tab:ListOfFields} 
\end{deluxetable*}

\section{Findings}
\label{sec:Findings}

Below we offer a detailed report for each of the ten sources reported
in B07.  Summarizing our results for the impatient reader we find four
of the eight single-epoch transients [RT\,19840502
(\S\ref{sec:RT19840502}), RT\,19860122 (\S\ref{sec:RT19860122}),
RT\,19840613 (\S\ref{sec:RT19840613}), RT\,19860115
(\S\ref{sec:RT19860115})] and one of the two-month transients
(RT\,19870422; \S\ref{sec:RT19870422}) to be artifacts. Our
re-analysis finds that the remaining long-duration transient,
RT\,20010331 (\S\ref{sec:RT20010331}) did not have a significant
detection when imaged with AIPS (SNR of 3.2) but the {\it Miriad}
imaging yields an SNR of 7.4.

For the remaining single epoch sources [RT\,19920826, RT\,19970205,
RT\,19970528, RT\,19990504] our analysis (undertaken by DAF; see
\S\ref{sec:ObsReana} for a summary of our data reduction) finds
reduced SNR.  The discrepancy with respect to the results reported
in B07 concerned us and so another author (GB) reanalyzed these
fields.  Some of the discrepancies arise from UV data flagging but
even when the same flagging is used different SNR algorithms yield
measures which differ by 1 to 1.5\,$\sigma$.  Since the sources lie
close to threshold of detection even a small shift of a single
$\sigma$ has an exponential effect in their confidence. In effect,
the reality of this group (taken as a whole) apparently depends on
details of algorithms in AIPS and {\it Miriad} -- the investigation
of which is beyond the scope of this paper. Below in
\S\ref{sec:RT19920826}--\ref{sec:RT19990504} we detail the analyses
and a summary of the SNRs of these sources can be found in
Table~\ref{tab:ListOfFields}.

We also re-investigated the lone transient candidate reported in
\cite{ofb+11}.  Our re-examination which now correctly includes the
number of independent beams that were searched for shows that there
is a four percent probability that this candidate is due to noise
(see \S\ref{sec:OtherSurveys}).  In the next section
(\S\ref{sec:RevisedB07}) we synthesize these findings and present
our conclusions about the B07 transients.

\subsection{RT\,19840502}
\label{sec:RT19840502}

\begin{figure}
\centerline{\includegraphics[angle=270,scale=0.4]{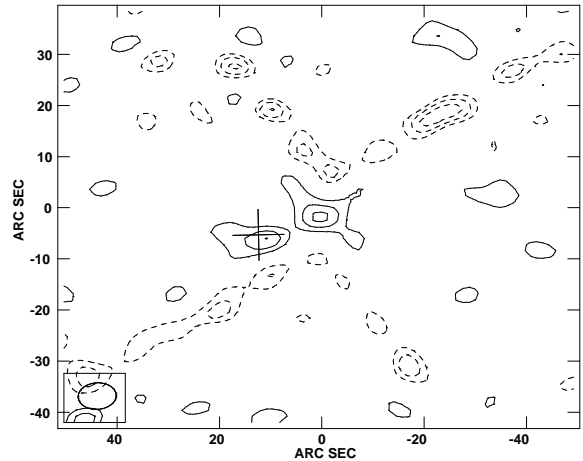}}
\caption{\small
  Image of RT\,19840502 marked by cross. The feature at the center (0,0)
  is an artifact (``phase center'').  Once the bad visibility
  data is removed the transient candidate is not visible on the
  final image. Contours are displayed in steps from $-1$, $-0.75$,
  $-$0.5, 0.5, 1.0, 1.5\,mJy, with negative (positive) contours
  given by dashed (sold) lines. The synthesized beam size is shown
  in the lower left.}
\label{fig:19840502} 
\end{figure}

B07 report finding a transient close to the pointing center
(13$^{\prime\prime}$) with a primary beam-corrected flux density of
$448\pm 74\,\mu$Jy, or a SNR=6.1. We imaged the calibrated data set
and confirmed the presence of emission at this level at the reported
position. However, we identified a phase center artifact in the
visibility data for the left hand polarization of the upper IF band.
The effect of this artifact in the image plane is to create strong
positive and negative slidelobes, with the positive feature identified
as a transient source (Figure~\ref{fig:19840502}).  When the upper IF
band data are removed the resulting image is noise-like and the peak
flux density at the nominal position of the reported transient is
$191\pm 97\,\mu$Jy. Additional observations were taken during the same
epoch at 15\,GHz. The peak flux density at the same position is
$-243\pm 200\,\mu$Jy.

\subsection{RT\,19840613}
\label{sec:RT19840613}

B07 report finding a transient coincident with a possible host
galaxy ($z=0.040$) and with a primary beam-corrected flux density
of $566\pm81\,\mu$Jy, or a SNR=7.0. Our deconvolved
image shows a source at that location.  Gaussian fitting suggests
that the source is resolved and this conclusion is supported with
an integrated flux density ($715\pm 218\,\mu$Jy) being clearly
larger than the estimate of the peak flux density of $388\pm
82\,\mu$Jy. Imaging the lower and upper IFs separately we find
another discrepancy.  The peak flux density in the lower IF is four
times weaker than the upper IF band.

\begin{figure}
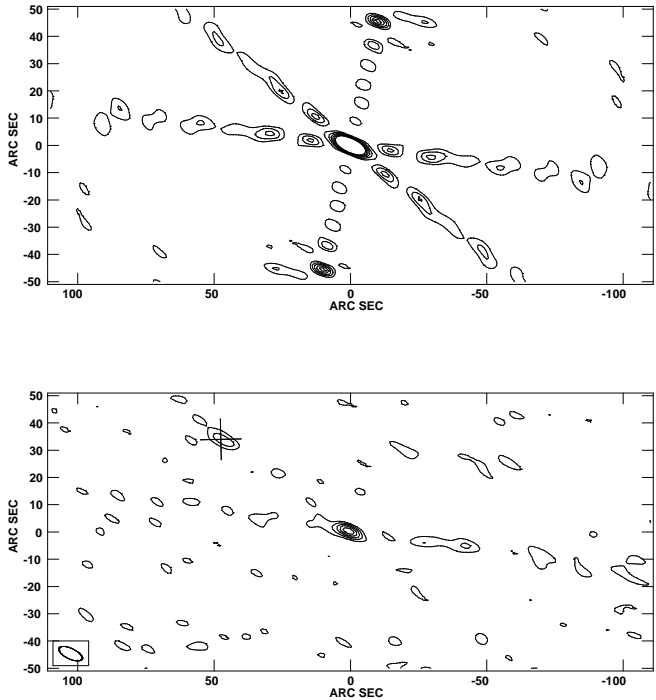

\includegraphics[angle=270,scale=0.35]{RT19840613_dbem.eps}\\
\includegraphics[angle=270,scale=0.35]{RT19840613_dim.eps}
\caption{\small
  {\it Top:} The dirty beam for the RT\,19840613 field.  This
  sub-image shows the northwestern side of the beam. The peak is
  located in the lower right hand corner. The contours are 5, 10,
  15, 20, 25 and 30\% of the peak flux. {\it Bottom:} The dirty
  image toward RT\,19840613. A cross marks the location of transient.
  Contours are 0.13, 0.23, 0.33, 0.43, 0.53, 0.63\,mJy\,beam$^{-1}$.
  Note that the transient candidate RT\,19840613 lies at the same
  angle and position as a side-lobe from the bright source
  J150123+781806 located in the bottom right corner of this image.
} 
\label{fig:19840613} 
\end{figure}

The likely source of the problem can be seen by comparing the dirty
image with the dirty beam (Figure \ref{fig:19840613}).  RT\,19840613
appears to be an uncleaned side-lobe of J150123+781806, one of the
brighter persistent sources detected in 452 images made by B07 at
5\,GHz. The putative transient is 56$^{\prime\prime}$ away from
J150123+781806 to the northwest, close to a local maximum (10\% of
peak) in the dirty beam at this location. This side-lobe artifact
is stronger in the upper band but it is still present in the lower
band.

Deconvolution does not fully remove the side-lobe from the image
and the effect is to produce a false transient. We investigated
whether the artifact is due to short-timescale ($\sim10$\,min)
variability of J150123+781806 but after dividing the visibility
data in half (by time) and re-imaging, we found no evidence for
variability.  The artifact may be due to some low level interference
picked up by some  antennas  baselines but we were not able to
identify the bad data. It is possible that RT\,19840613 is a real
transient that unfortunately lies on the side-lobe of the dirty
beam, but its Gaussian fit parameters and its variation in the lower
and upper sidebands suggest that it is not a real source.

Additional observations were made during this epoch at 1.5\,GHz.
The data quality is good and the images have no obvious artifacts.
No source is visible at the transient position. The peak flux density
is $133\pm113\,\mu$Jy.

\subsection{RT\,19860115}
\label{sec:RT19860115}

\begin{figure}
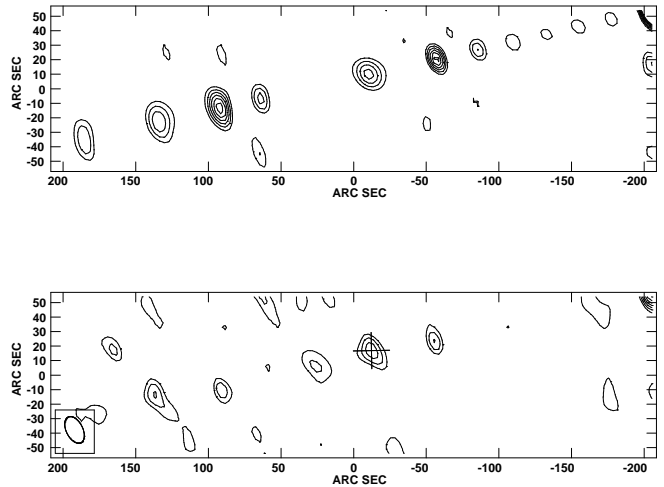

\includegraphics[angle=270,scale=0.35]{RT19860115_dbem.eps}\\
\includegraphics[angle=270,scale=0.35]{RT19860115_dim.eps}
\caption{\small
  {\it Top:} The dirty beam for the RT\,19860115 field.  This
  sub-image shows the southwestern side of the beam. The peak of the
  beam is located in the upper right hand corner. The contours are
  7.5, 12.5, 17.5, and further in steps of 5\% up to 37.5\% of the
  peak flux.  {\it Bottom:} Dirty image toward RT\,19860115. A
  cross marks the location of the transient. The contours are 0.15,
  0.25, 0.35, 0.45, 0.55, 0.65 and 0.75\,mJy\,beam$^{-1}$. Note that
  the transient candidate RT\,19860115 lies at the same angle and
  position as a side-lobe from the bright source J150123+781806
  located in the upper right corner of this image.
}
\label{fig:19860115} 
\end{figure}

B07 report a transient with a primary beam-corrected flux density of
$370\pm 67\,\mu$Jy, or a SNR=5.5. Some of the same issues with the
image of the previously discussed RT\,19840613 were also seen for
RT\,19860115. In Figure~\ref{fig:19860115} we show the dirty image
along with the synthesized beam.  The RT\,19860115 appears to be an
uncleaned side-lobe of J150123+781806 and hence not a real transient.
RT\,19860115 lies at the same angular distance (3.3$^\prime$) and
position angle (100$^\circ$ CCW CC) of a local maximum in the
side-lobe structure (25\% of the peak beam).  We were unable to
identify the source of these strong side-lobes.  As in the case of
RT\,19840613, we were able to rule out that the strong side-lobes
originated from short-term variability of J150123+781806.

\subsection{RT\,19860122}
\label{sec:RT19860122}

B07 report a transient with a primary beam-corrected flux density of
$1586\pm 248\,\mu$Jy, or a SNR=6.4. Images made with all the
visibility data do show emission at the position of RT\,19860122 with
the reported peak flux density. However, there does appear to be some
erroneous visibility data resulting in low-level rings in the image
centered at the phase center. The bad data were traced to the lower
IF. If the data for the lower IF band are removed there is no source
at the reported transient position (peak flux density of $303\pm
189\,\mu$Jy).

\subsection{RT\,19870422}
\label{sec:RT19870422}

This is one of two transients found by binning individual epochs into
two-month averages. B07 report a 5\,GHz transient with a primary
beam-corrected flux density of $505\pm 83\,\mu$Jy, or a SNR of 6.1.
The source J150050+780945.5 is positionally coincident with a blue
host galaxy ($z=0.249$) and has been suggested by \cite{np11} as a
candidate afterglow from a binary coalescence event. Three single
epochs were used to form the average. They are in YYYYMMDD format:
19870414, 19870422 and 19870429. A close inspection of the data
reveals a problem.  The last two of the three epochs used to find
RT\,19870422 were pointed to a different part of the sky, an error
that was made when the original observing files were written.  The
source is actually J210133.67+373528, and appears only as a transient
when added inappropriately to the data from the standard on-the-sky
test field. In hindsight we recognize these to be the same type of
errors as seen earlier \citep[see][]{obg+10}.

\subsection{RT\,20010331}
\label{sec:RT20010331}

This is the second of the transients which were found by binning
individual epochs into two-month averages. B07 report a 5\,GHz
transient with a primary beam-corrected flux density of $697\pm
94\,\mu$Jy, or a SNR=7.4.  Separately, \cite{ctb11}
reported a marginal X-ray source at this position.

Eight single epochs were used to form the average. They are in
YYYYMMDD format: 20010306, 20010314, 20010321, 20010328, 20010403,
20010411, 20010418 and 20010425. We calibrated and imaged the eight
epochs of observations from 2001 March 6 to 2001 April 25 (all in
B-configuration).  While the four quiescent sources from B07 can be
seen in this deep image, we do not identify a significant source at
the position of RT\,20010331. The brightest peak within the
synthesized beam (natural weighting) is $42\pm 13\,\mu$Jy (or
3.2$\sigma$). An independent reduction using the AIPS package by GB
confirms the absence of this source.

The first problem lies with the primary beam corrections reported
in B07.  RT\,20010331 lies 4.4$^{\prime}$ from the phase center,
close to the 10\% response radius of the primary beam.  Our estimate
for the rms noise of $13\,\mu$Jy lies within a few percent of the
theoretical value.  The beam-corrected rms noise in this case (at
10\% response of the primary beam) would then be about $129\,\mu$Jy,
not the $94\,\mu$Jy given in B07.  Rather than correcting at the
10\% radius, it appears that the flux density and noise in B07 were
mistakenly corrected at the 14\% power level. This multiplicative
error has no impact on the signal-to-noise.

In order to investigate this signal-to-noise discrepancy between our
image and B07, we split the data in various ways (separate epochs,
months of March and April epochs, adjacent IF bands) and re-imaged,
looking for a bright peak. None were found.

The discrepancy between B07 and the work reported apparently can be
traced back to differences in the {\it Miriad} and AIPS imaging
packages. Our calibrated visibility data, when processed through {\it
  Miriad} using nearly identical imaging parameters as those in AIPS,
gives a flux density of $91\pm13\,\mu$Jy (7$\sigma$). We have no
explanation for the discrepancy between the two results obtained from
AIPS and {\it Miriad}.  It is worth noting that the peak flux
densities of the persistent sources identified by B07 agree in these
images to $0.5\sigma$. For this paper we accept the analysis given
here.

There was one epoch (2001 March 6) in which data were also taken at
5\,GHz. The peak flux density at the position of RT\,20010331 is
$-27\pm 39\,\mu$Jy. In summary, we find no evidence to support that
RT\,20010331 is a significant detection.

\subsection{Remaining Four Single Epoch Sources}
\label{sec:RemainingFourTransients}

\subsubsection{RT\,19920826}
\label{sec:RT19920826}

B07 report a transient with a primary beam-corrected flux density of
$642\pm101\,\mu$Jy, or a SNR=6.4. We confirm a source
at this position but with a slightly reduced SNR. Using natural
weighting of the gridded visibilities, our measured peak flux density
is $460\pm 80\,\mu$Jy, or 5.8$\sigma$. Despite the lower significance,
there is some confidence that RT\,19920826 is real since it appears in
both IFs with comparable flux densities.  Further investigation by GB
shows that the difference in the SNR between B07 and the analysis here
can be traced to differences in flagging of the UV data (of two
specific antennas).
 

\subsubsection{RT\,19970528} 
\label{sec:RT19970528}

B07 report a transient with a primary beam-corrected flux density
of $1731\pm 232\,\mu$Jy, or a SNR=7.5. These
observations were taken during a time when some antennas were being
moved to the B configuration and so we applied antenna position
corrections (via AIPS task \texttt{VLANT}) before calibration.

RT\,19970528 is 6.8$^\prime$ from the center of the image, where
the response of the antennas is only 16\% of their peak. The
uncorrected flux density is $270\pm 47$. A point source search (AIPS
task \texttt{SAD}) of the four million pixels enclosed interior to
a radius around this candidate, shows six other un-cataloged
candidates with similar signal-to-noise. The primary-beam corrected
flux density $1.7\pm 0.3\,$mJy is similar that from to B07. This
estimate does not include a correction for temporal smearing due
to the sources distance from the phase center, nor for the added
uncertainty in the magnitude of the primary beam correction beyond
the 20\% point.  Correcting for these terms we get $2.8\pm 0.5\,$mJy.

GB re-investigated this source and found that with the B07 analysis
the SNR varies with the approach (peak/rms, \texttt{SAD}, \texttt{JMFIT})
between 6.2, 7.4 and 7.1.  (without any corrections discussed above).
The corresponding SNRs for the image discussed above is 6.1, 5.1
and 6.8.  The source is a bit ``ratty'' and this may explain the
variation in SNR. We therefore find  this source to be a weak
detection.

Additional observations were made during this epoch at 8.5\,GHz.
The source lies close to the first null ($6.4\pm 0.3^{\prime}$ from
the phase center) of the beam at this frequency. The attenuation
by the primary beam is severe and hence the sensitivity is not
sufficient to provide any strong spectral index constraints.


\subsubsection{RT\,19990504}
\label{sec:RT19990504}

B07 report a transient with a primary beam-corrected flux density
of $7042\pm 963\,\mu$Jy, or a SNR=7.3. 
\cite{ctb11} report an X-ray source in the vicinity of the radio
source. Deep multi-wavelength data is consistent with the X-ray
source arising from a QSO but located five arcseconds from the
putative radio transient. 



We find the
(uncorrected) peak flux density is $290\pm 51\,\mu$Jy (or an SNR
of 5.7).
RT\,19990504 lies 8.9$^{\prime}$ from the phase center, close to
the first null ($11.1\pm 0.6^{\prime}$ from the phase center) where
we would not expect to find sources. The polynomial expressions
used in AIPS to correct for the beam attenuation are increasingly
inaccurate outside the 20\% response radius, and they are not
applicable close to the null. The azimuthally averaged measured
value of the beam response is 1.8\%, implying a flux density
correction factor of 55$\times$ \citep{cp10} or $16.0\pm 2.8\,$mJy.

As with RT\,19970528, the analysis of B07 data by GB finds SNR of
5.6 (peak/rms), 7.3 (\texttt{SAD}) and 6.8 (\texttt{JMFIT}) and a
similar variation with the image reported here.  It is worth noting
that the source may be extended and also that a visual inspection
of the annular region reveal a number of similar sources.  We
conclude that RT\,19990504 is not a robust detection.

Additional observations were made during this epoch at 8.5\,GHz.
An image was made from these higher frequency data but, as expected,
no source was found.

%


%

\subsubsection{RT\,19970205}
\label{sec:RT19970205}

This is the only single-epoch transient identified at 8.5\,GHz by B07.
They report a primary beam-corrected flux density of $2234\pm
288\,\mu$Jy, or a SNR=7.8.  RT\,19970205 is 4.4$^\prime$ from the
center of the image, where the response of the antennas is only 9.8\%
of their peak. There is an elongated source at this position with a
peak flux density of $231\pm 41\,\mu$Jy (using natural weighting).
There is some indication that the source is extended since the
integrated flux density is about twice the peak flux density.



The peak flux density was likely underestimated by B07 since they did
not fully correct for bandwidth or temporal smearing effects.  These
data were taken in the BnA array with an integration time of 3 1/3 s
and a 50-MHz bandwidth. In this observing configuration, the effects
of temporal smearing with this dump time will likely reduce the peak
flux density by only a few percent. However, chromatic aberration is
expected to be larger with the effect of smearing a point source along
the radial direction \citep{psb89}.


Accounting for the bandwidth effect, we measure a peak flux density
of $3.4\pm 0.6\,$mJy (SNR=5.7). Next, due to the uncertainty about
the form of the primary response function beyond the 20\% point
(AIPS task PBCOR), we only know that the peak flux of RT\,19970205
lies in the range of 2.9--4.4\,mJy.  As noted above the source is
likely extended and this may account for the variation in SNR (from
$\approx 6$ to 7.6).


\subsection{Other Surveys}
\label{sec:OtherSurveys}

\cite{ofb+11} presented a survey for 5\,GHz radio transients at low
Galactic latitudes.  Most of the data were reduced in near-real
time (2\,hr delay) and transient candidates were followed up with
radio, visible light and X-ray instruments.  The authors reported
a single transient candidate, J$213622.04+415920.3$ with a peak
specific flux of $2.36\pm 0.41\,$mJy (5.8\,$\sigma$) and with no
obvious optical counterpart.

We have re-analyzed this data set and confirm that the SNR is 5.8.
Considering that the search resulted from inspecting $1.1\times 10^7$
independent beams (\S\ref{sec:NumberIndependentBeams}) and assuming
Gaussian statistics, we find that the probability of the highest event
found in these many independent beams is attributable to chance or
noise is 3.6\% (see \S\ref{sec:A}). Therefore, we advocate that this
candidate is not a real event.

\section{The Revised B07 Transient Rate}
\label{sec:RevisedB07}

As noted in \S\ref{sec:Introduction} the ten transients reported
in B07 were remarkable for their apparent abundance (1.5 per square
degree, at any given epoch) {\it and} the lack of quiescent optical
counterparts.  In \S\ref{sec:Findings} we found that five of the
B07 transients are  artifacts arising out of (rare) data acquisition
problems or imaging artifacts (see Table~\ref{tab:ListOfFields}).
We find that the 2-month transient RT\,20010331 is not well detected.
We argue that the single transient reported by \cite{ofb+11} is
also not a robust detection.

In \S\ref{sec:Findings} we present our analysis for the remaining four
transients (all of which are single-epoch transients): RT\,19920826,
RT\,19970205, RT\,19970508 and RT\,19990504. We find SNRs which are
typically 1-$\sigma$ below that reported in B07. The revised and B07
SNRs are given in Table~\ref{tab:ListOfFields}. We note that an
independent pipeline written in ParselTongue \citep{kett06}, a Python
interface to AIPS, and run on three of these four sources confirms the
lower SNR values found here \citep{bell11}: RT\, 19920826 (SNR=6.0),
RT\, 19970528 (SNR=4.4), and RT\, 19990504 (SNR=4.0).

\begin{figure}[htbp] 
    \centering
    \includegraphics[width=2.7in]{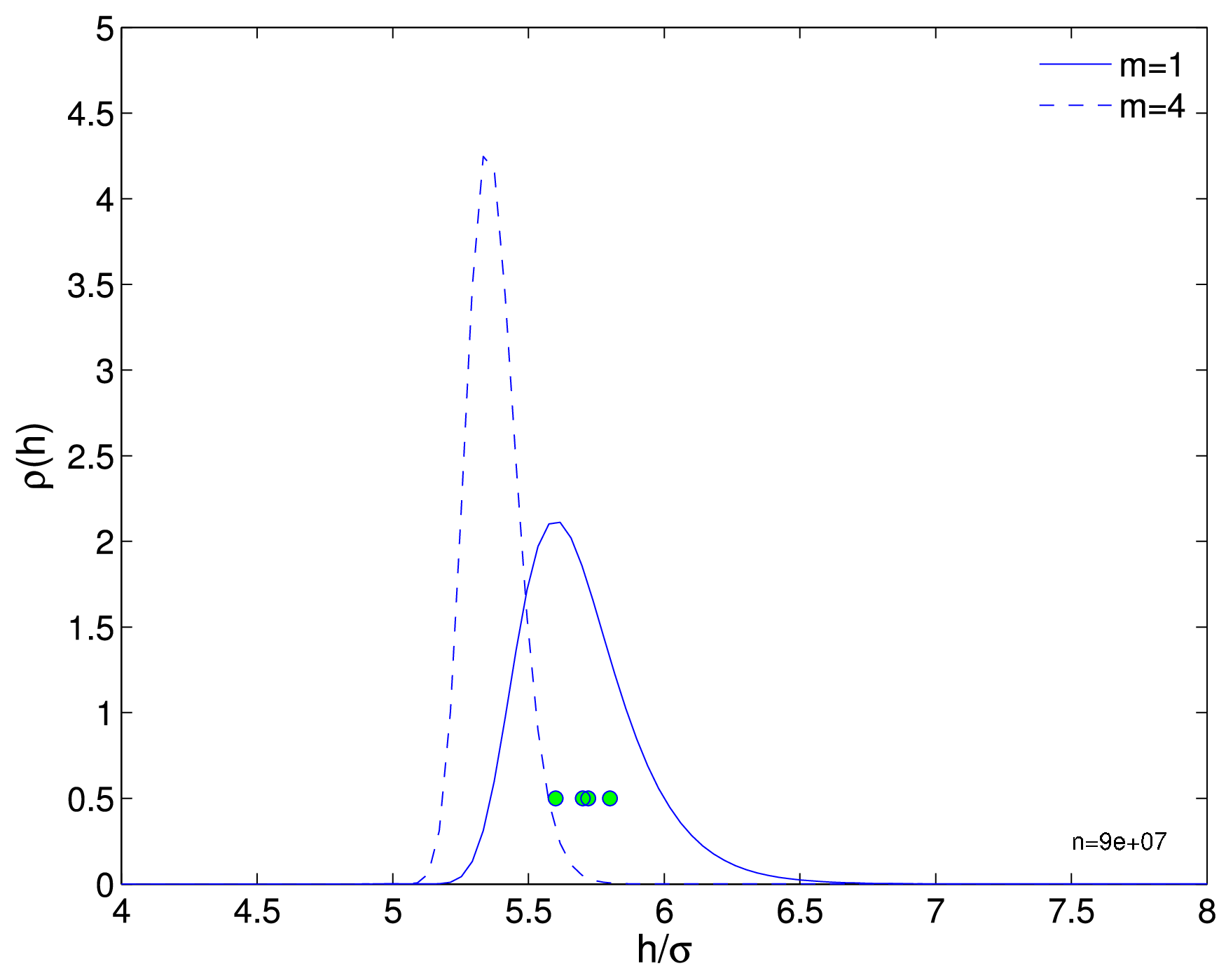}
    \includegraphics[width=2.7in]{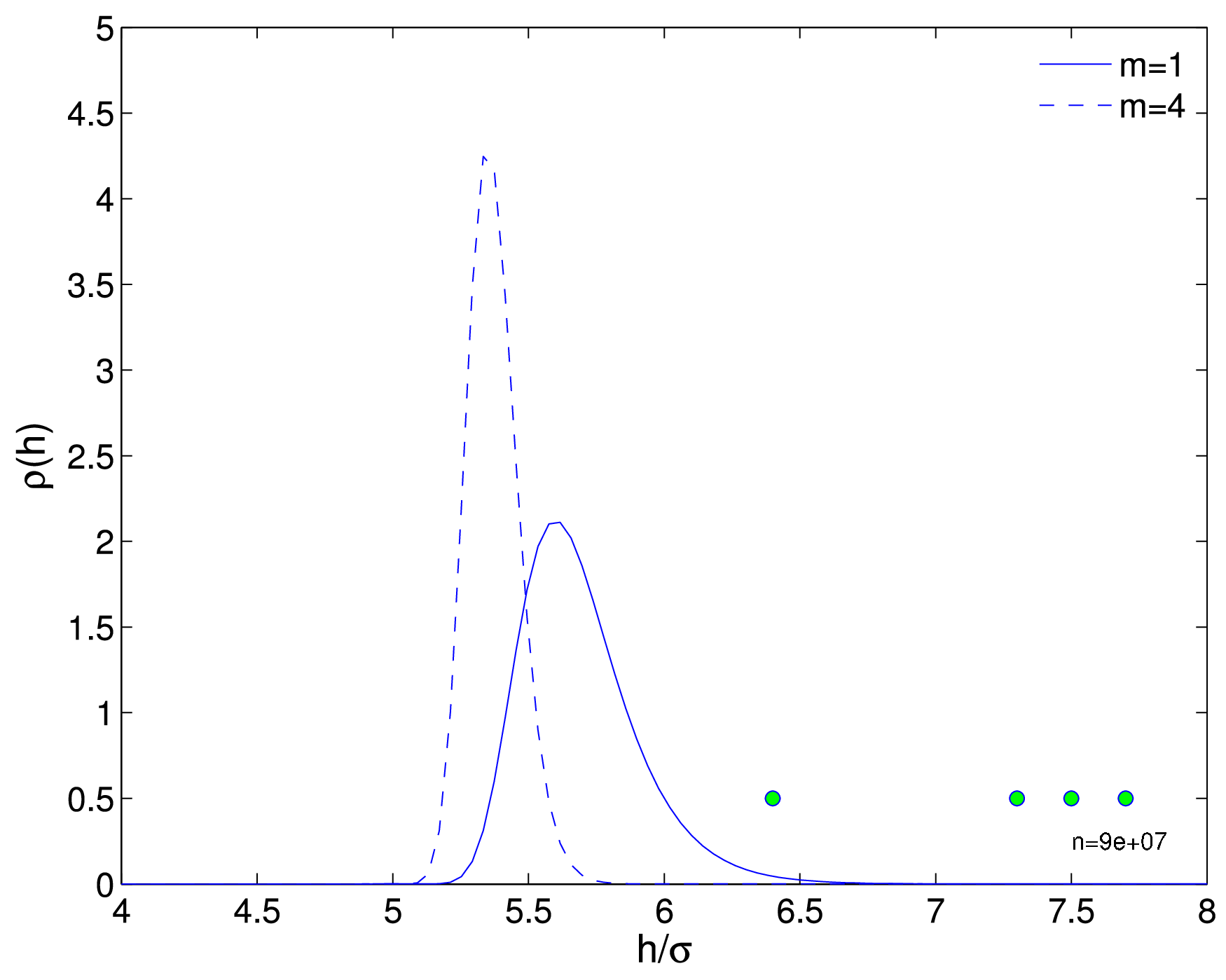}
    \caption{\small
      The probability density function of the highest value ($m=1$)
      and the fourth highest value ($m=4$) of a population of
      $n=9\times 10^7$ Gaussianly distributed random numbers with
      zero mean and unit variance. The probability density function
      plotted here is discussed in \S\ref{sec:A} and the justification
      for $n$ can be found in \S\ref{sec:NumberIndependentBeams}.
      The dots represent the SNRs of the following single-epoch
      transients discussed in Table~\ref{tab:ListOfFields}.  (from
      left to right): RT\,19905054, RT\,19920826, RT\,19970205 and
      RT\,19970228. (Top): SNRs as reported in this paper. (Bottom):
      SNRs as reported in B07. Note that the vertical location of
      the dots is arbitrary.  For details of these density distributions
      see \S\ref{sec:A}.
}
\label{fig:PdistPoints}
\end{figure}

The lowering of SNR (from between 7 and 8 to between 5 and 6) has a
pernicious effect when the number of independent beams which were
search is included. In \S\ref{sec:NumberIndependentBeams} we estimate
this number to be $n\approx 9\times 10^7$.  In \S\ref{sec:A} we derive
the probability density function for the highest $m$ values of $n$
Gaussian random numbers\footnote{Theory informs us that the statistics
  of beam values or equivalently pixel values of interferometric maps
  should follow Gaussian distribution.}. In
Figure~\ref{fig:PdistPoints} we plot the density function for the
highest value ($m=1$) and the fourth highest value ($m=4$).  As can be
seen from this Figure~\ref{fig:PdistPoints}, if the SNRs reported here
are accepted then the global case for the remaining B07 transients is
entirely weakened.  If, on the other hand, the SNRs reported in B07
are accepted then the four transients reported in B07 do argue for a
new class of radio transients.

The above approach of using a fixed threshold for all epochs does
not result in optimal detection. In particular, the threshold for
a low resolution survey is lower than that for a higher resolution
survey (since the latter has a correspondingly larger number of
synthesized beams).  B07 addressed this problem by requiring that
the probability of a false detection (PFD) in an individual epoch
was constant and less than $N$ where $N$ is the total number of
images. With this approach, the expectation number of false detections
is 1 for the entire survey.  Applying the B07 method we find the
following PFDs:
RT\,19920826 (log(PFD)=$-5.02$);
RT\,19970205  ($-2.74$);
RT\,19970528  ($-2.77$);
RT\,19990504  ($-4.61$).
With this more refined approach only RT\,19920826 and 19990504
survive.  However, for reasons discussed in \S\ref{sec:RT19990504}
we have misgivings about RT\,1990504.

An entirely different approach\footnote{This test was recommended to
  us by J. Condon.} (and in some ways orthogonal to the above SNR
based approach) is to look at the angular distribution of the
transient sources with respect to the primary axis\footnote{We assume
  that all antennas are pointed in the same direction and this
  direction is both the pointing axis as well as the phase center.}.
Basic interferometry theory informs us that the dirty image is simply
the Fourier transform of the visibility data. As such the radiometric
noise in the dirty image should be independent of the angular offset
from the phase center.  In contrast, the point source sensitivity
decreases as one goes away from the pointing center and this is
governed by the primary beam response (assuming that the spectral
resolution of the survey is high enough that the delay beam is larger
than the primary beam). Thus, once the minimum SNR for detection is
fixed, cosmic sources should be concentrated towards the pointing
direction whereas noise spikes (masquerading as threshold point
sources) should be uniformly distributed.

\begin{figure}[htbp] 
   \centering
   \includegraphics[width=2.7in]{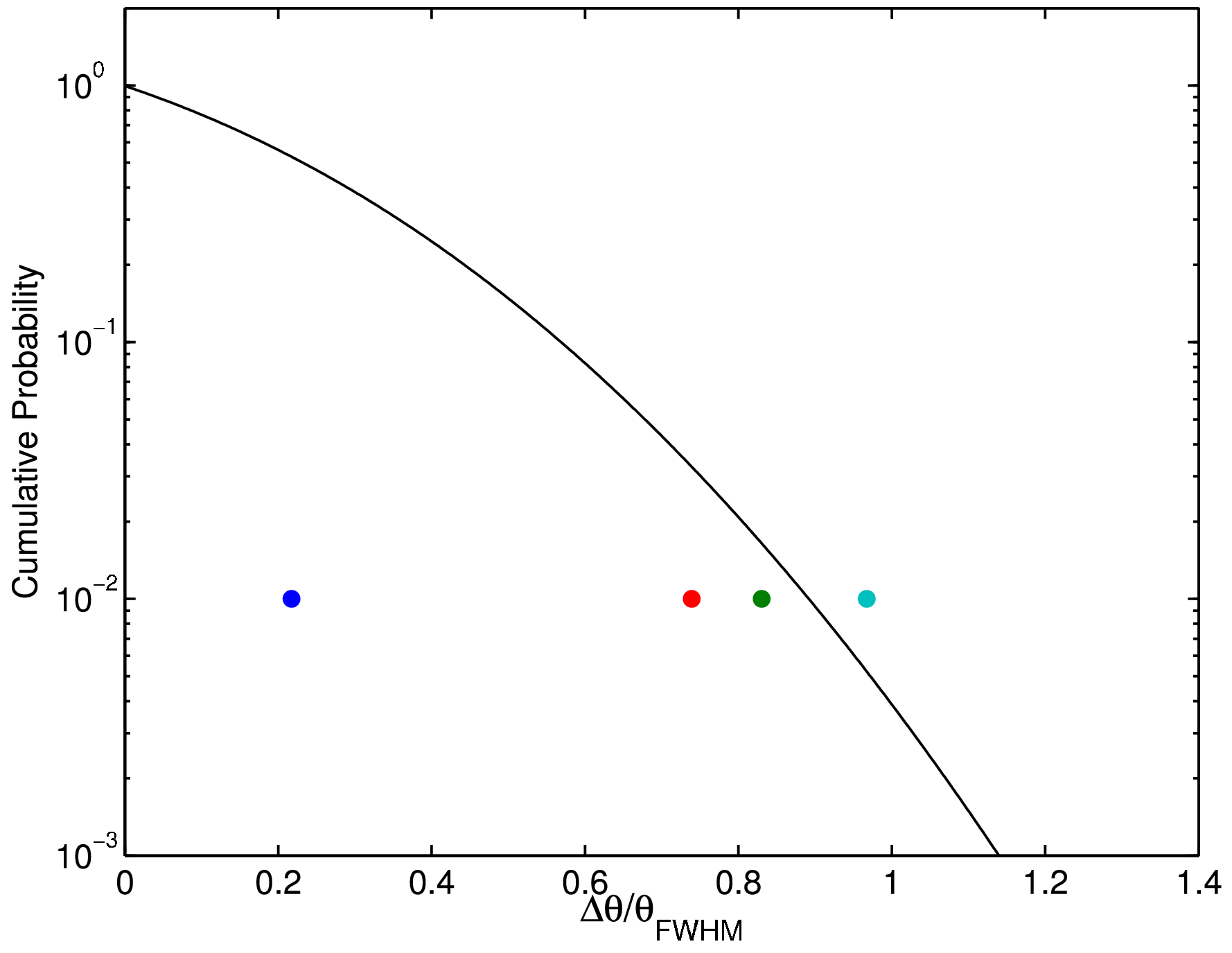}
   \caption{\small
     The cumulative probability of finding a cosmic source (but
     integrated from angular offset of infinity to zero) as a function
     of the angular offset with respect to the pointing center. The
     angular offset is normalized in units of $\theta_{\rm FWHM}$.
     The sources are represented by dots and are (from left to right):
     RT\,19920826, RT\,19970528, RT\,19970205, and RT\,19990504. The
     points are deliberately placed at cumulative probability of 1\%.
     RT\,19920826 is firmly within the region where one naturally
     expects cosmic sources.}
   \label{fig:PrimaryBeamPoints}
\end{figure}

In \S\ref{sec:PrimaryBeam} we derive the expected distribution of
cosmic sources as a function of the angular offset. In
Figure~\ref{fig:PrimaryBeamPoints} we plot the expected cumulative
distribution and also the angular offset of the four sources which are
not artifacts but whose SNR seems to be under dispute, namely
RT\,19920826 (\S\ref{sec:RT19920826}), RT\,19970205
(\S\ref{sec:RT19970205}), RT\,19970528 (\S\ref{sec:RT19970528}) and
RT\,19990504 (\S\ref{sec:RT19990504}). From this Figure one can see
that only RT\,19920826 lies in the expected region whereas the
remaining three are collectively improbable.

In summary, two different statistical tests, one based on SNR and the
other making use of the spatial signature provided by the primary
beam, suggest that of the remaining four sources detected at
threshold, only one, namely, RT\,19920826 is a good detection.  Thus a
simple interpretation of our re-analysis is that the rate of B07
transients is considerably lower than that reported by B07, perhaps an
order of magnitude smaller.

We acknowledge that the discrepancy between the analysis presented
in B07 and the analysis presented here (and re-investigated) is disturbing.
In the previous section (\S\ref{sec:RemainingFourTransients}) we
investigate the reasons for the discrepancy in the SNRs and variously
find possible and plausible causes: flagging of data, the choice
of data reduction package (AIPS versus {\it Miriad}) and the specific
method used to compute SNRs. However, none of these explanations
are satisfactory.  We are continuing this investigation but at the
present time we consider this topic to beyond the scope of the
paper.

In contrast, the sources which we find as artifacts have ready
explanations (see Table~\ref{tab:ListOfFields}). One source is a
result of the file header containing a pointing direction of the
previous pointing. Another is due to a systematic associated with
local signals (RFI). These signals do not have the natural fringe rate
of cosmic sources and appear as candidates close to the phase center.
Two are side-lobes of a stronger sources.\footnote{Unfortunately,
  side-lobes are the exception to the expectation of Gaussian
  statistics for interferometric images. It is said that ``the Central
  Limit theorem covers a large number of sins but not all sins.''}
Suffice to say that such rare errors will be found if one inspects
sufficiently large number of beams!

\begin{figure*}
  \centering
  \includegraphics[angle=0,scale=0.8]{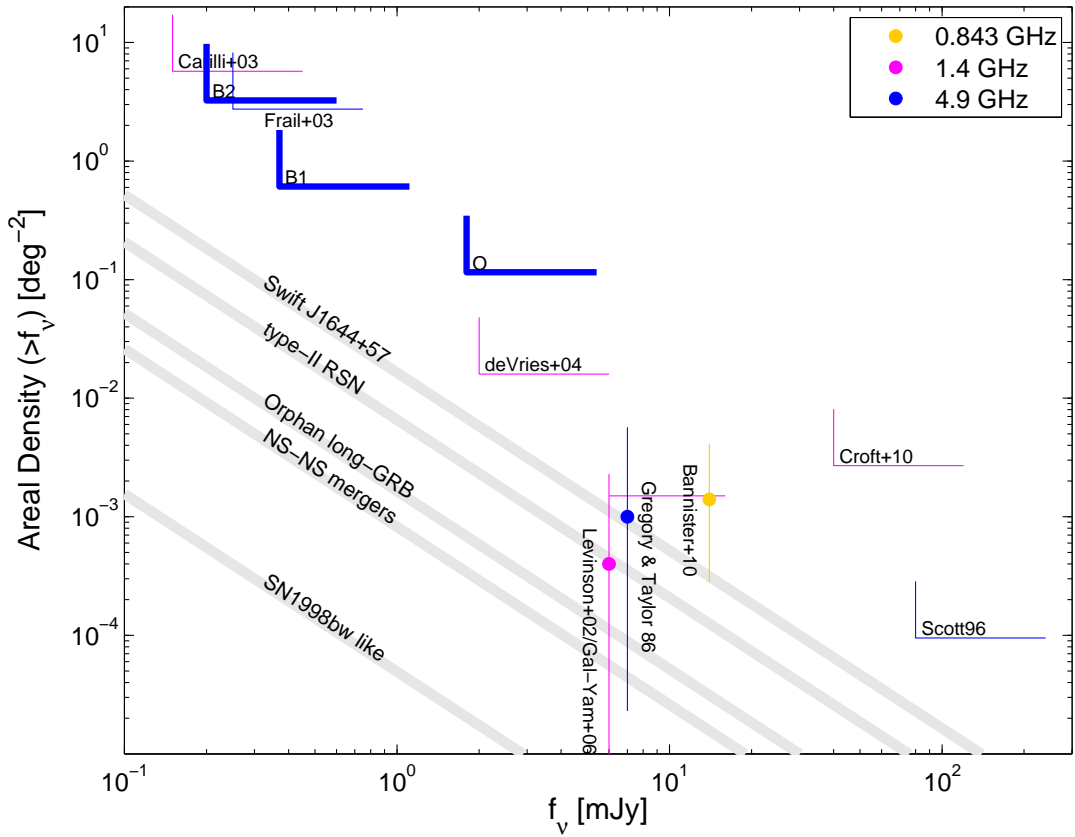}
\caption{\small
  Cumulative areal density of transients as a function of peak flux
  density for all major transient surveys. Most of the surveys are
  upper limits and the allowed phase space is above and to the right
  of the L-shaped symbol. The three dark blue L's (annotated as B2
  for the 2-month transients from B07, B1 for the single-epoch
  transients from B07 and O for the lone transient reported in
  \citealt{ofb+11}) are the upper limits derived as a result of the
  analysis presented here. These limits were derived by assuming no
  detection (whence a Poisson upper limit of 3 at the 95\% confidence level;
  see \S\ref{sec:UpperLimitPoisson}) and survey areas summarized in
  \cite{obg+10}.  
} 
\label{fig:rates} 
\end{figure*}

\section{A Revised Look at the Transient Radio Sky}\label{sec:logNlogS}

In Figure~\ref{fig:rates} we plot the areal densities of three known
transients (SN1998bw-like, type-II RSN and Swift J1644+57-like) and
that of two expected classes (NS-NS mergers and orphan afterglows
of long duration GRBs) of transients.  The areal density of these
five classes is also summarized in Table~\ref{tab:ListOfTrans}. We
briefly discuss each of these five classes of transients below.

The traditional extra-galactic radio transients are type II radio
supernovae.  The rate that we present here is based on the single
radio SN detected in a blind radio survey by \citet{gop+06} and it
agrees with an independent estimate by \citet{lcfk11}. As illustrated
by the example presented in \cite{gop+06}, the main advantage in the
search for radio SNe in a blind survey is the unique view that it
provides to the otherwise hidden population of heavily obscured SNe.
The observed rate of other types of radio SNe (e.g.  ordinary type
Ib/c SNe) are considerably lower than that of type II supernovae and
are not discussed further.

Radio emission is expected from both classical long duration gamma-ray
bursts (GRBs) as well as the more abundant but less luminous GRBs
exemplified by GRB\,980425 associated with the energetic SN Ic
supernova, SN\,1998bw \citep{gvp+98,kfw+98}.  The radio emission
is far brighter than that of ordinary core collapse SN (II, Ib, Ic)
and the increased volume makes up for the intrinsically smaller
birth rate.  The recent discovery of the energetic supernova SN2009bb
\citep{scp+10} demonstrates that radio surveys can find such sources
without resorting to high energy (gamma-ray) missions.

\citet{lowg02} estimated the number of afterglows from classical GRBs
and whose explosion axis is directed away from us (``orphan''
afterglows).  The expected rate depends strongly on the poorly
constrained $\gamma$-ray beaming. On one hand this makes any rate
prediction uncertain. On the other hand, even a non-detection by the
kind of survey that we discuss below will provide an independent and
unique constraint on the average opening angle of long gamma-ray
bursts, their true rate and total energy output
\citep{rpd+08,npg02,tp02}.  The areal density in
Figure~\ref{fig:rates} is derived using a typical opening angle of 10
degrees.

A surprising and apparently an important development in the field of
radio transients took place just this year with the discovery of a
radio transient associated with the nucleus of a modest size galaxy.
The source, Swift\,J1644+57 was initially detected as a hard X-ray
transient \citep{bkg+11}. Subsequent follow up found a bright, compact
self-absorbed radio counterpart, localized at the center of a normal
galaxy at $z=0.354$ \citep{zbs+11,ltc+11}. The current view is that
Swift\,J1644+57 arose from a relativistic jet produced when a star was
tidally disrupted as it passed too close to an otherwise dormant
super-massive nuclear black hole \citep[e.g.][]{bgm+11}.  Shortly,
thereafter, a second candidate non-thermal tidal disruption event
(TDE) was recently proposed \citep{ckh+11}. Events such as these give
us an opportunity to study the activity of 10$^7$-10$^8$ M$_\odot$
supermassive black holes in otherwise normal galaxies.

The areal density in Figure~\ref{fig:rates} is calculated assuming an
observed rate of 0.2 ${\rm yr^{-1}}$ Swift\,J1644+57-like events and a
gamma-ray beaming factor of $10^{3}$ \citep{zbs+11,bgm+11}.
Nominally, Swift\,J1644+57-like sources appear to be the most frequent
extra-galactic transients that will be found in radio transient
searches. We acknowledge that the uncertainty of both the observed
rate and the gamma-ray beaming is high and the true rate may be
significantly different.\footnote{Estimates based on theoretically
  predicted TDE rates and luminosities \citep{gm11,bow11,vkf11} result
  in areal densities that vary by three orders of magnitude.  The rate
  that we predict here is consistent with the upper range of these
  predictions.}

Now we come to the most uncertain as well as potentially the most
important extra-galactic radio transient -- the merger of two neutron
stars (or a black hole and a neutron star). It is generally accepted
(or expected) that short hard bursts are on-axis explosions of these
mergers \citep{nakar07,mb11}. As in long duration GRBs, radio
emission is expected by afterglow (on-axis or orphan). The rates
are highly uncertain because there are very few observations of
short hard GRBs.  Thus there still continues to be a debate about
the geometry of these explosions (``jetted'' or not).  Next, while
the expected radio emission is straightforward to estimate (subject
to the usual parametric uncertainties of the energy fractions of
relativistic electrons and magnetic field) an additional uncertainty
is the density of the ambient gas (which is necessary for the
production of the afterglow emission).

Regardless of the uncertainty whether neutron star mergers are the
sources of short GRBs or not, a substantial sub- and mildly
relativistic outflow is expected to be ejected during the merger.
\cite{np11} estimate radio emission from these outflows. The areal
density in Figure~\ref{fig:rates} is calculated based on their
estimates\footnote{The rate density of such mergers is poorly
  constrained.  It ranges between 10 to $10^4\,{\rm
    Gpc^{-3}\,yr^{-1}}$ for NS-NS mergers
  \citep{p91,nps91,kkl+04,aaa+10}}, assuming a NS-NS merger rate of
300 ${\rm Gpc}^{-3} {\rm yr}^{-1}$ and that any merger ejects
  $10^{50}$ erg of a mildly relativistic outflow. We note that
  \cite{np11} suggested that RT\,19870422 was the radio emission from
  the remains of a neutron star merger. However, as noted in
  \S\ref{sec:Findings} this source is an artifact.


\section{Way forward: New Surveys}\label{sec:FutureSurveys}

There are sound reasons to continue the exploration of the dynamic
radio sky. Radio searches are an ideal way to discover core-collapse
supernovae embedded in or behind dusty regions. The discovery of
SN\,2009bb shows that large radio searches can find urgently needed
additional examples of nearby low luminosity GRBs. Next, the many
rewards of radio follow up observations of Swift\,1644+57 (accurate
localization, energetics, beaming, outflow velocity) show the
tremendous diagnostic power of radio observations of this entirely new
class of extragalactic transients.

As exciting as these developments are, the search for new classes of
radio transients has involved several false starts. The euphoria that
followed the discovery of a highly dispersed (and therefore argued to
be of extragalactic origin) millisecond burst \citep{lbm+07} was
rapidly diminished by the discovery of many such bursts, presumably of
terrestrial origin \citep{bbe+11}; but see \cite{kkl+11}.  Similarly,
a long-duration transient found by \cite{lowg02} and \cite{gop+06} was
later traced to a glitch in the VLA on-line data taking system
\citep{obg+10}, apparently affecting 0.29\% of all FIRST survey
pointings \citep{thwb11}.  Finally, our re-analysis (see
\S\ref{sec:RT19870422}) shows that the claim of late time radio
emission from neutron star coalescence \cite{np11} is premature.

\begin{figure}[hbt]
\includegraphics[angle=0,scale=0.45]{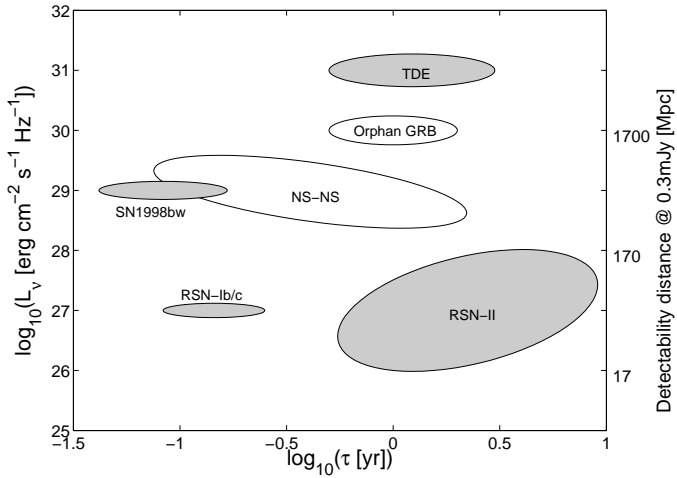}
\caption{Phase space diagram showing the predicted radio luminosity
  versus evolutionary time scales for several types of long duration
  radio transient populations. Transparent zones indicate source
  populations which are typically optically thin, while grey zones
  indicate source populations that are expected to be optically thick
  before maximum light, evolving to an optically thin phase at later
  times. A similar optical version of this figure can be found in
  \citet{rkl+09} and a more comprehensive figure which includes short
  duration events is in SKA Memo 97.
}
\label{fig:phase} 
\end{figure}

Here we focused on the potentially new class of long duration radio
transients reported in B07. We  rule out six of the ten transients
and cast doubts on some of the remaining ones. However, even with
one or two survivors the long duration transients of B07 remain of
great interest. First, even with a diminished number of transients
the implied areal rate of B07 transients would be comparable to the
recently established class of Swift\,J1644+57 transients. However,
{\it unlike any other long duration radio transient} (to wit,
supernovae; active stars; tidal disruption events; gamma-ray burst
afterglows, beamed or otherwise) the B07 transients are remarkable
for the absence of a quiescent optical counterpart.  

The event RT\,19920826 survives two independent tests. As such it
useful to speculate on the origin of this transient. The absence of an
optical (B07) and near IR quiescent source could mean one of two
origins. The event is extra-galactic in origin but the host galaxy is
faint enough not to have been detected (as does happen for a few
percent of long duration GRB host galaxies) or that it is offset from
a host galaxy (as in the case for a few short hard bursts;
\citealt[][]{berg09}). Alternatively, the event is a Galactic neutron
star and we have to then appeal to an optically invisible Galactic
(neutron star) population \citep{obg+10}.


Clearly, a new survey which can net a dozen of such sources (but
brighter!)  would help resolve the origin. For the Galactic hypothesis
one expect no quiescent counterpart even at HST sensitivity whereas
detectable galaxies, in the majority, are expected for the
extra-galactic hypothesis.

There are great gains in the discovery of new classes of radio
transients but at the same time the path to true success is littered
with false starts or easy speculations. The way forward must
incorporate the lessons learnt from the false starts.  We elaborate on
this conclusion below.

To start with we believe that the field of radio transients (at
least in the decimeter band) is sufficiently mature that any new
survey which just sets an upper limit relative to the known population
is of marginal value.  Future surveys have to be sufficiently deep
and cover large enough sky that success (i.e.  detection of a few
to many transients) is assured.  In our opinion, this means that a
survey should be designed to find at least one or more Swift\,J1644+57-like
transient (Figure~\ref{fig:rates}).

Next, timely and multi-wavelength follow-up is essential. For example,
the transients reported in \cite{gop+06}, \cite{gt86} and
\cite{bmg+11a,bmg+11b} have plausible origins as supernovae and
Swift\,J1644+57-like sources (see Figure~\ref{fig:rates}). However,
the lack of timely follow-up or deep multi-wavelength followup of
these events preclude us from coming to a definitive conclusion.

Finally, the search should be restricted to sources with a high level
of significance. This certainly means paying attention to the large
number of beams searched. However, at low thresholds and with large
number of beams (cf \S\ref{sec:Findings}) it would be prudent to set
thresholds beyond mere statistical considerations\footnote{Separately
  we caution that the discussion of statistics assumes that the
  underlying statistics are Gaussian to a very high degree of
  precision. As noted in the text, the sidelobes of strong sources add
  an additional source of non-Gaussian noise.}. A threshold of 9 or
even $10\sigma$ may be appropriate.  Alternatively, an immediate
verification of a transient by deeper observation or a confirmation by
observations at other wavelengths would allow detection of transients
closer to threshold.

We start with a discussion of two recent developments.  \cite{bfs+11}
undertook an ambitious program similar in spirit to B07, namely the
investigation of fields surrounding VLA calibrators.  Sources with
duration between 4 and 40\,d were searched for.  The total integration
time was 435\,hr. No transient source in the GHz range with flux
greater than 8\,mJy was found. The authors place an upper limit to the
areal density of 0.032\,deg$^{-2}$.  Assuming $S^{-3/2}$ scaling this
areal density is 4.4\,deg$^{-2}$ and is well above the B07 rate.

The FIRST survey imaged the sky in an hexagonal grid, in which each
position in the survey footprint was observed on average 3--4 times
\citep{bwh95}.  \cite{thb+11} used this fact to construct light
curves of sources detected in individual FIRST survey snapshots.
They identified 1627 variable candidates with variability exceeding
$5\sigma$.  This effort is probably the largest variable and transient
survey ever carried out.  One disadvantage of such a survey for
transient identification is that the co-added  images are not much
deeper than a single epoch image.  This make it hard to tell if an
apparent transient source is really a transient or just a variable
source that exceeded the detection threshold in one of the epochs.

The limitations discussed above lead us to suggest a new EVLA survey
specifically tailored to systematically explore the radio sky.
Following that we review a far more ambitious survey -- the VAST
key project on ASKAP.  For the discussion below we will adopt the
rates summarized in Table~\ref{tab:ListOfTrans}. The rates are
specified to a flux density of 0.3\,mJy and are extrapolated to higher
flux densities as N($>S)\propto S^{-3/2}$.

\subsection{EVLA Survey}
\label{sec:EVLASurvey}

A moderately ambitious survey with the EVLA can result in great
progress.  This survey has two virtues.  One, the EVLA offers
excellent spatial resolution.  Next, the EVLA is fully commissioned
and is working to specifications.

Specifically consider a 100 square degree survey undertaken in the
2--4\,GHz band. An integration time of 85\,s results in a
sensitivity of 0.3\,mJy (10-$\sigma$).  A single epoch covering 100
square degrees would require 50 hours. As can be seen from
Figure~\ref{fig:phase} (taken from Table~\ref{tab:ListOfTrans})
such a survey would have to explore a variety of time scales to
probe the emerging classes of transient. Fifteen epochs could
reasonably cover the range of a week to years.

Noting the great importance of multi-wavelength imaging data, such a
survey would sensibly focus on regions of sky where considerable
multi-wavelength data (including radio) exists. One such region is,
for example, the SDSS equatorial stripe \citep{hbw+11}.  Furthermore,
the high instantaneous sensitivity of the EVLA makes rapid follow up
of newly transients possible.

After the first four epochs the reference images will be twice as deep
as the survey field. A single new epoch would then yield about ten
Swift\,J1644+57-like sources and four supernovae. Ten such images may
find a new example of an SN\,1998bw-like event, several clear examples
of orphan afterglows and have an excellent chance of finding the first
examples of neutron star mergers.  We note that these different
classes of objects have different characteristics, both in duration
(Figure~\ref{fig:phase}) and also in host magnitudes and location with
respect to host galaxy.  Therefore, it is possible to distinguish
between different types of objects.

As note earlier, rapid verification of a transient (either by additional and
deeper radio observations or observations at other wavelengths) can
reduce the requirement for a high detection threshold. This would then
require close rapid reduction -- well within the reach of modern computers.

\subsection{VAST (ASKAP)}

The Variable \&\ Slow Transient (VAST) is an approved key project of
ASKAP\footnote{http://www.physics.usyd.edu.au/sifa/vast/index.php}.
The VAST-Wide survey aims to survey in the 1.2\,GHz band about 10,000
square degrees every day for 2 years. With 40-s integrations the
expected single-epoch rms is 0.5\,mJy (VAST Memo\#1).  Since the
survey is planned to be undertaken daily the reference image will be
built up quite rapidly.

A comparison of data obtained at a new epoch when compared with the
built-up reference image can be expected to detect about six
Swift\,J1644+57-like sources.  Type II SN and Swift\,J1644+57-like
sources are relatively long lived and so one could consider averaging
the daily images (to say 10 days). The resulting summed data has a
sensitivity of 0.16\,mJy.  The expected number of transients per
such summed image is nominally 32 (Type II SN), 82 (Swift\,J1644+57),
8 (orphan afterglows) and 4 (neutron star coalescences).

To sum up,  the dynamic radio sky remains a rich area for exploration.
Based on what we know about the areal rates for the {\it known}
transient sources, future synoptic radio imaging surveys are expected
to yield substantial numbers of exotic transients. Such surveys
will also provide the definitive test for the B07 population.


\acknowledgments

DAF thanks Jim Condon and Alicia Soderberg for important discussions
early on in this project.  We thank Steve Croft for a most careful
reading of the paper and J. Condon for making several insightful
suggestions.

The National Radio Astronomy Observatory is a facility of the
National Science Foundation operated under cooperative agreement
by Associated Universities, Inc. SRK thanks
the Department of Astronomy, University of Wisconsin at Madison for
their hospitality. EOO is supported by an Einstein fellowship and
NASA grants.  SRK's research in part is supported by NASA and NSF.
This research has made use of data from the University of Michigan
Radio Astronomy Observatory which has been supported by the University
of Michigan and by a series of grants from the National Science
Foundation, most recently AST-0607523.  This research has made use
of NASA's Astrophysics Data System.

\appendix
\section{A. Probability Density Function of $m$th maximum}
\label{sec:A}

In this section, our goal is to compute the probability density
function of the $m$th highest value of $h_j, j=1,2,3...n$. Let $p(h)$
be the probability density function of $h_j$ with
$P(h)=\int_{-\infty}^hp(h)dh$ being the cumulative function. We denote
the $m^{\rm th}$ highest value by $H_m$. Thus the maximum of the
series of measurements is $H_1$ and the minimum value is $H_n$.

Let $\rho(H_m)$ be the probability density function of $h=H_m$.  This
means that at least one of the measurements lies is in the range
$[H_m,H_m+dH_m]$.  The probability density for this event is $p(H_m)$.
Next, then $n-m$ measurements must lie below this range and $m-1$
above this range.  The probability for any value to be smaller than
$H_m$ is $P(H_m)$ and the probability for a value to higher than $H_m$
is is $1-P(H_m)$.  This is now a binomial distribution with $n-1$
total values. Thus the probability density function for $H_m$ is
        \begin{equation}
        \rho(H_m) = np(H_m)\times \frac{(n-1)!}{(n-m)!(m-1)!}P(H_m)^{n-m}\Big[1-P(H_m)\Big]^{m-1}.\
        \label{eq:rho}
        \end{equation}
The first combinatorial factor of $n$ accounts for the possibility
that $H_m$ can occupy any position in the sequence. The second
combinatorial factor, $^{n-1}C_{m-1}$ account for the combinations
satisfying the condition that $n-m$ values lie below $H_m$ and $m-1$
lie above $H_m$.  For both the maximum ($m=1$) and minimum ($m=n$)
Equation~\ref{eq:rho} simplifies to that expected from basic
considerations.

Now let us consider the specific case where $h$ follows Gaussian statistics:
        \begin{equation}
                p(h) = \frac{1}{\sqrt{2\pi}}\exp(-h^2/2)
        \end{equation}
where $h$ is normalized in units of $\sigma$.   
The probability that an event is extreme or lies within the range $\pm h$ is 
        \begin{eqnarray}
                \phi(h) &=&\int_{-h}^{+h} p(h)dh =
                \frac{2}{\sqrt{\pi}}\int_0^{h\sqrt{2}}\exp(-z^2)dz= {\rm erf}(h/\sqrt{2}).
        \end{eqnarray}
The probability that an event is extreme in only one sense, maximum or minimum,
and lies outside the range $[-\infty, h]$ (say) is thus
        \begin{eqnarray}
                P(h) &=& \frac{1}{2}\Big[{1+\rm erf}\Big(\frac{h}{\sqrt{2}}\Big)\Big]= 
                1-\frac{1}{2}{\rm erfc}\Big(\frac{h}{\sqrt{2}}\Big)
        \end{eqnarray}
where  ${\rm erfc}(x)=1-{\rm erf}(x)$.

Consider the case where $n\gg 1$ (say $10^6$ or more) and $m$ is
small, say 10. Then we can approximate $n-1\approx n$, $n-2\approx
n$, ..., $n-m+1\approx n$. Furthermore using the approximation,
$(1-x/n)^n \approx \exp(-x)$, we find in the limiting case where
$H_m$ is greater than a few [so that $1-P(H_m)\ll1$]:
        \begin{eqnarray}
        \rho(H_m)&=&\frac{n}{(m-1)!}p(H_m)\exp\Big[-\frac{(n-m)}{2}{\rm erfc}\Big(\frac{H_m}{\sqrt{2}}\Big)\Big]
        \Big[\frac{n}{2}{\rm erfc}\Big(\frac{H_m}{\sqrt{2}}\Big)\Big]^{m-1}.
        \end{eqnarray}

\section{B. Distribution of Sources within the Primary Beam}
\label{sec:PrimaryBeam}

Provided that there is sufficient spectral resolution, the response of
an interferometer to a source follows the antenna response function
(``primary beam'').\footnote{We assume that the phase center coincides
  with the pointing axis of the primary antenna.} We will assume that
the response function is azimuthally symmetric and specified by
$g(\theta)$ where $\theta$ is the angular offset from the pointing
axis.

Let the areal density of sources with flux density greater than
$S$ be a power law, say,
$N(>S)\propto S^\alpha$. For Euclidean geometry
and most reasonable luminosity functions, 
$\alpha=-3/2$. Next, we note that detection is really finding
sources at a given SNR and above. Fortunately the noise
distribution for an interferometric image is uniform. 
Thus 
a source with a given flux density will have an SNR, $\mathcal{S}$, that
scales with the primary beam response,
$\mathcal{S}\propto Sg(\theta)$. 
The number of sources above a certain SNR and contained
outside an angular radius of $\theta_0$ is 
        \begin{equation}
                n(>\mathcal{S}; >\theta_0) \propto \int_{\theta_0}^{\infty} 2\pi\theta\Big(\frac{\mathcal{S}}{g(\theta)}\Big)^\alpha d\theta
        \label{eq:nS}
        \end{equation}
        
\begin{figure}[htbp] 
    \centering
    \includegraphics[width=3in]{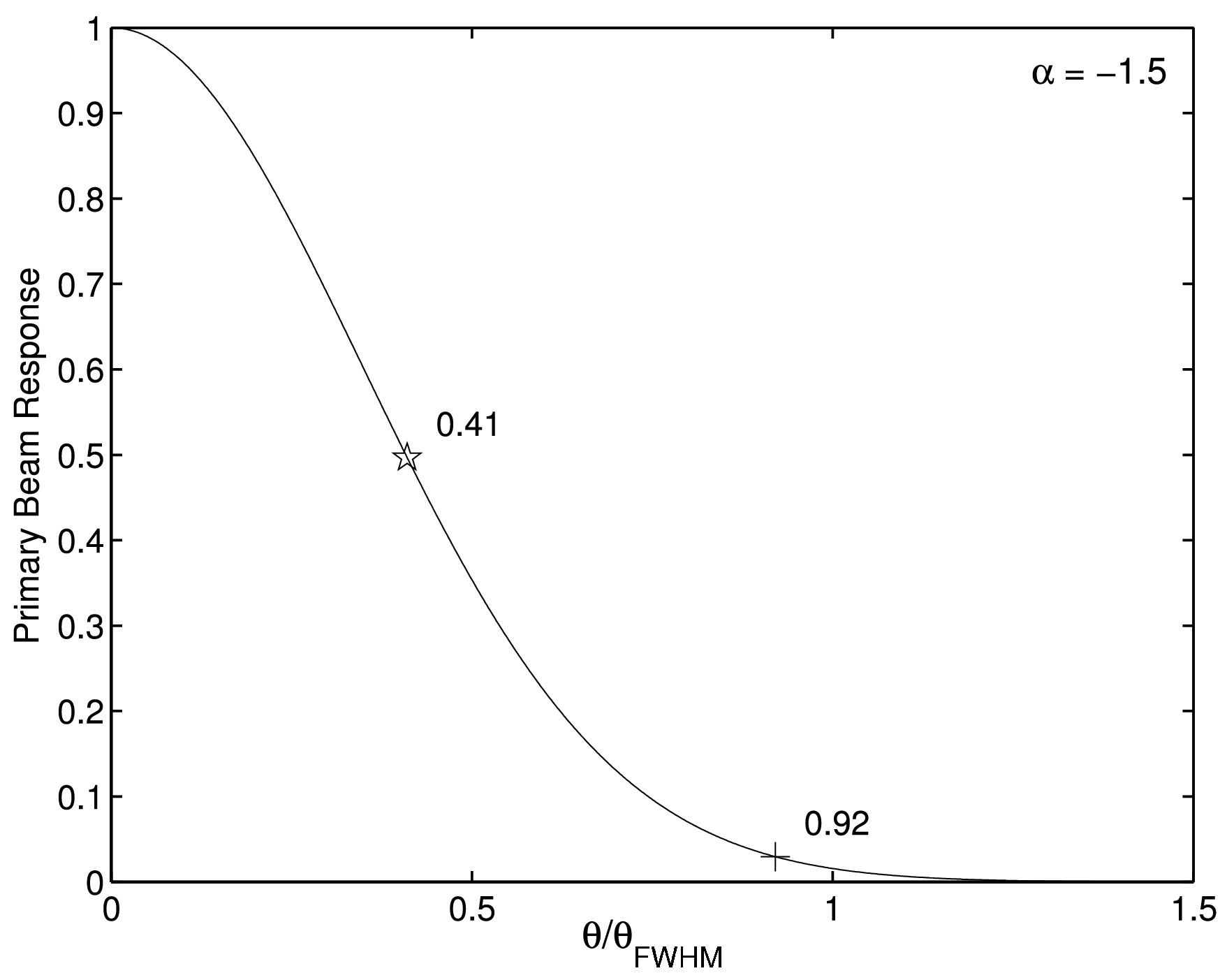} 
    \caption{\small
    The expected distribution of sources above a threshold SNR as a function
    of $\theta/\theta_{\rm FWHM}$ from the pointing aixs and assuming $\alpha=-3/2$. The
    response curve is normalized by insisting that the integral of the curve (from $\theta=$
    to $\theta=\infty$) is unity. 
    Fifty percent of the sources are within $0.41\theta_{\rm FWHM}$ (marked by a pentagram) 
    and 97\%
    within $0.92\theta_{\rm FWHM}$ (marked by a cross; at this radius the primary beam gain is only
    0.1 relative to that on-axis).
    }
    \label{fig:PrimaryBeamDistribution}
 \end{figure}

For the specific case of a Gaussian beam, 
        \begin{equation}
        g(\theta)\propto \exp\Big[-\frac{1}{2}(\theta/\theta_*)^2\Big]
        \label{eq:Gaussian}
        \end{equation}
where the traditional ``full width at half maximum'' (FWHM) is
$\theta_{\rm FWHM}=\sqrt{\ln(256)}\theta_*$. Substituting Equation~\ref{eq:Gaussian}
into Equation~\ref{eq:nS} we obtain
        \begin{equation}
        n(>\mathcal{S};>\theta_0)\propto \Big(\mathcal{S}/g(\theta_0)\Big)^\alpha
        \end{equation}
Half the sources will be detected outside the radius 
$\theta_h=\sqrt{-\ln(4)/\alpha}\,\theta_*$. For $\alpha=-3/2$ we obtain
$\theta_h=\sqrt{\ln(16)/3}\,\theta_*\approx 0.97\theta_*$. 
The expression $\theta_h=\sqrt{1/6}\,\theta_{\rm FWHM}\approx 0.4\,\theta_{\rm FWHM}$
is more useful. A plot of $n(>\mathcal{S};>\theta)$ can be found
in Figure~\ref{fig:PrimaryBeamDistribution}.
 
\section{C. Upper Limit for a Non Detection (Poisson)}
\label{sec:UpperLimitPoisson}

It is not unusual to find no source after undertaking a survey. We
wish to determine an upper limit to the number of sources that were
being searched.  The number detected is given by Poisson statistics.
The probability of finding $r$ sources is then given by
        \begin{equation}
                p(r) = \frac{{\lambda}^r}{r!}\exp(-\lambda)
        \label{eq:PoissonDistribution}
        \end{equation}
where $\lambda$ is the Poisson parameter and equal to $\langle r\rangle$. 

Our goal is to determine the maximum value of $\lambda$ given a
non-detection.  As the value of $\lambda$ is increased the probability
of detection, by which we mean the probability of detecting one or
more events, also increases.  This probability is $p(1)+p(2)+...$
which we note is $1-p(0)$.  This probability  can be set to the
desired confidence level, $\mathcal{P}$ and thence
        \begin{equation}
                \mathcal{P}=1-p(0)=1-\exp(-\lambda).
        \end{equation}
The reader with a stronger physical bent may find the complement,
$p(0)=1-\mathcal{P}$,  more appealing.  Regardless, we find $\lambda=
[3,4.6,6.9]$  at a confidence level of [95\%, 99\%, 99.9\%].

\section{D. Number of Independent Beams}
\label{sec:NumberIndependentBeams}

Here we compute $n$, the number of independent beams for the VLA
data that went into the analysis of B07 and \cite{ofb+11}.  For
B07, a circular region with a radius of two times the half-power
radius was searched for each epoch.  The number of independent beams
per epoch is the ratio of that area to the area of the synthesized
beam in that particular epoch.  For individual epochs, this value
ranged from as small as $10^3$ to as large as $10^6$.  The total
numbers of independent beams for the 5 and 8.4 GHz data are $9.3
\times 10^7$ and $4.5 \times 10^7$, respectively.  The smaller
number of independent beams for 8.4\,GHz are due to the tapering
of the visibility data, increasing the typical synthesized beam
size at 8.4\,GHz.

For \cite{ofb+11} the search was made for transients in 4.65\arcmin
radius circular region. The FWHM of the synthesized beam is
$\approx4$\,arc-seconds.  Though the goal was 16 epochs per pointing
we only achieved an average of 15.7 images. With 141 epochs we
derive $n=1.1\times 10^7$.

\end{document}